\documentclass[12pt]{article}
\usepackage{amssymb,epsfig}
\newcommand{\es}{$\mathrm{E_6}$}

\begin{document}

\hspace*{\fill}UH-511-968-00

\hspace*{\fill}WUE-ITP-2000-018

\hspace*{\fill}hep-ph/0007310

\vspace{1cm}

\begin{center}
\textbf{\Large Displaced Vertices in Extended Supersymmetric Models}

\vspace{10mm}
S.~Hesselbach\footnote{e-mail: hesselb@phys.hawaii.edu} 

Department of Physics and Astronomy, University of Hawaii, Honolulu,
HI 96822, USA

\vspace{5mm}
F.~Franke\footnote{e-mail: fabian@physik.uni-wuerzburg.de}, 
H.~Fraas\footnote{e-mail: fraas@physik.uni-wuerzburg.de}

Institut f\"ur Theoretische Physik, Universit\"at W\"urzburg, D-97074
W\"urzburg, Germany
\end{center}

\vspace{5mm}
\begin{abstract}
In extended supersymmetric models with additional singlet Higgs fields
displaced vertices could be observed if the decay width of the 
next-to-lightest supersymmetric particle
becomes very small due to a singlino dominated LSP.
We study the supersymmetric parameter space where displaced vertices 
of the second lightest 
neutralino exist in the NMSSM and an \es\ inspired model.
For a mass difference between LSP and NLSP of more than 10~GeV
the singlet vacuum expectation value has to be at least of the order of
$100$~TeV in order to obtain a lightest neutralino with
a singlino component large enough for displaced vertices.
\end{abstract}

\section{Introduction}

Displaced vertices are often assumed to be a characteristic 
feature of the neutralino sector in extended supersymmetric models
with gauge singlets. 
Especially in the Next-to-Minimal Supersymmetric Standard Model
(NMSSM) with a singlino-like lightest supersymmetric particle
(LSP) displaced
vertices are expected in a large part of the parameter space
\cite{hugonie}.
They may appear if the mixing character of the lightest
neutralino 
$\tilde{\chi}^0_1$ -- which is also the LSP -- 
is dominated by the singlino component so that its couplings 
to the other supersymmetric particles are strongly suppressed.
Then in models with conserved $R$-parity all supersymmetric
particles first decay into the next-to-lightest SUSY particle (NLSP)
which is assumed to be the second lightest neutralino  
$\tilde{\chi}^0_2$. 
The NLSP
finally decays into the LSP with a 
decay vertex displaced from the production
vertex
if the decay width is small enough.

The direct production of singlino dominated neutralinos
in the NMSSM and the \es\
model at an $e^+e^-$ linear collider with polarized beams 
has been discussed in \cite{singlinoprod}.
Slow decays in models containing additional singlet superfields with
intermediate-scale vacuum expectation values have been considered in
\cite{spmartin}.
The aim of this study is a detailed analysis of the
low energy parameter regions 
where displaced vertices of the $\tilde{\chi}^0_2$ 
exist in 
two representative extended supersymmetric models,
the NMSSM and an \es\ model with an additional neutral
gauge boson. These models are shortly described
in Sec.~2.

For this purpose we first discuss in Sec.~3    
scenarios where the lightest neutralino has a large
singlino component.
The vacuum expectation value $x$ of the singlet field turns
out to be the crucial parameter responsible for a
singlino dominated $\tilde{\chi}^0_1$ leading to a 
displaced $\tilde{\chi}^0_2$ decay vertex.
Therefore we present in Sec.~4 the $\tilde{\chi}^0_2$ decay width in
representative scenarios as a function of $x$ and show the
parameter region where displaced vertices exist. 
Our main conclusion is that, depending on
the neutralino masses and the mass difference between
the lightest neutralinos, the singlet vacuum expectation value has to be
in the range between several TeV and $10^4$~TeV
in order to observe displaced vertices in these extended
models.

\section{NMSSM and E$_\mathbf{6}$ model}

The NMSSM is the minimal extension of the Minimal Supersymmetric 
Standard Model (MSSM) by a singlet Higgs
field $S$ with vacuum expectation value $x$ \cite{nmssm1}.
The masses and couplings of the five neutralinos 
depend on the the gaugino 
mass parameters $M_2$ and $M_1$, the ratio of the vacuum 
expectation values of the doublet Higgs fields ($H_1$, $H_2$) 
$\tan\beta = v_2/v_1$ as in the MSSM, and on 
the singlet vacuum expectation value $x$ and the 
trilinear couplings $\lambda$ and $\kappa$ in the superpotential
\cite{nmssm2}
\begin{equation}
W \supset \lambda H_1 H_2 S - \frac{1}{3} \kappa S^3 \,.
\end{equation}
In this paper we always assume the GUT relation $M_1/M_2= 5/3
\tan^2 \theta_W$.
For large $x \gg |M_2|$ a singlino dominated neutralino decouples in 
the neutralino mixing matrix and could become very light
for small values of the parameter $\kappa$.
Then the heavier neutralinos have MSSM character with
$\mu = \lambda x$.
Such light or even massless singlino dominated neutralinos are not
excluded by LEP \cite{franke}.

Since also light neutral singlet-like 
Higgs bosons can generally exist in the NMSSM
\cite{frankehiggs} the NMSSM Higgs sector may play
an important role for the decay characteristics of the neutralinos.
It contains five physical neutral Higgs bosons, three Higgs scalars and
two pseudoscalars whose masses and mixings depend on the soft scalar masses
$A_\lambda$ and $A_\kappa$ in addition to $\tan\beta$, $x$, $\lambda$ and
$\kappa$.

The \es\ model with an additional $\mathrm{U(1)'}$ factor in the low
energy gauge 
group and therefore a new neutral gauge boson $Z'$ is
a further extension of the MSSM beyond the NMSSM. It can be
motivated by superstring theory \cite{hr} and implies
an \es\ group as effective GUT group, which is broken to
an effective low energy gauge group of rank~5.
We consider an \es\ model 
with one additional singlet superfield 
in the Higgs sector \cite{e6model}.
To respect the experimental mass bounds for new gauge bosons
the singlet vacuum expectation value $x$ must be
larger than about 1.5~TeV~\cite{abe}.
The extended neutralino sector in this model contains six
neutralinos being mixtures of photino, zino, $Z'$ gaugino,
doublet higgsinos and singlino
\cite{e6model,e6neutralino,suematsu}. The $6 \times 6$ neutralino mass matrix
depends on six parameters: the $\rm SU(2)_L$, U(1)$_Y$ and
$\rm U(1)'$ gaugino
mass parameters $M_2$, $M_1$ and $M'$, $\tan\beta$, $x$
and the trilinear coupling $\lambda$ in the superpotential
\begin{equation}
W \supset \lambda H_1 H_2 S \,.
\end{equation}
In the \es\ model 
the $4\times 4$ submatrix of the MSSM-like neutralinos and the
$2\times 2$ submatrix of the exotic ones approximately decouple
because of the large values of $x$.
Then for very large values of $|M'| \gg x$ light singlino-like
neutralinos occur because of a see-saw-like \cite{seesaw}
mechanism in the $2\times
2$ submatrix \cite{decarlosespinosa}.

\section{Scenarios}
Since displaced vertices are only expected for an LSP with a
large singlino component we discuss in the following 
scenarios with a singlino-dominated $\tilde{\chi}_1^0$. 

In the NMSSM the singlino content of $\tilde{\chi}_1^0$
is described by the squared matrix element $|N_{15}|^2$
of the unitary $5 \times 5$ matrix $N$
which diagonalizes the neutralino mass matrix 
in the basis
$(\tilde{\gamma},\tilde{Z},\tilde{H_1},\tilde{H_2},
\tilde{S})$ of the photino, zino, the two doublet higgsinos and the
singlino. 
The MSSM content of $\tilde{\chi}_1^0$, described by $1 - |N_{15}|^2$, 
is crucial 
for the couplings of the heavier neutralinos to the LSP and especially 
for the decay width of the second lightest neutralino.

Similarly in the \es\ model the $6 \times 6$ neutralino mass matrix is 
diagonalized by an unitary $6 \times 6$ matrix $N$
with $|N_{16}|^2$ describing the singlino 
content of the LSP.
Here for $|M'| \gg x$
the lightest neutralino can be a nearly
pure singlino.
Since the very heavy $\tilde{Z}'$ with mass ${\cal O}(M')$ decouples
nearly completely
\cite{lc97,hesselb}, 
$1 - |N_{16}|^2$
describes the MSSM content of the LSP in very good
approximation.

\begin{figure}[ht]
\begin{picture}(16,8.6)
\put(0,.5){\epsfig{file=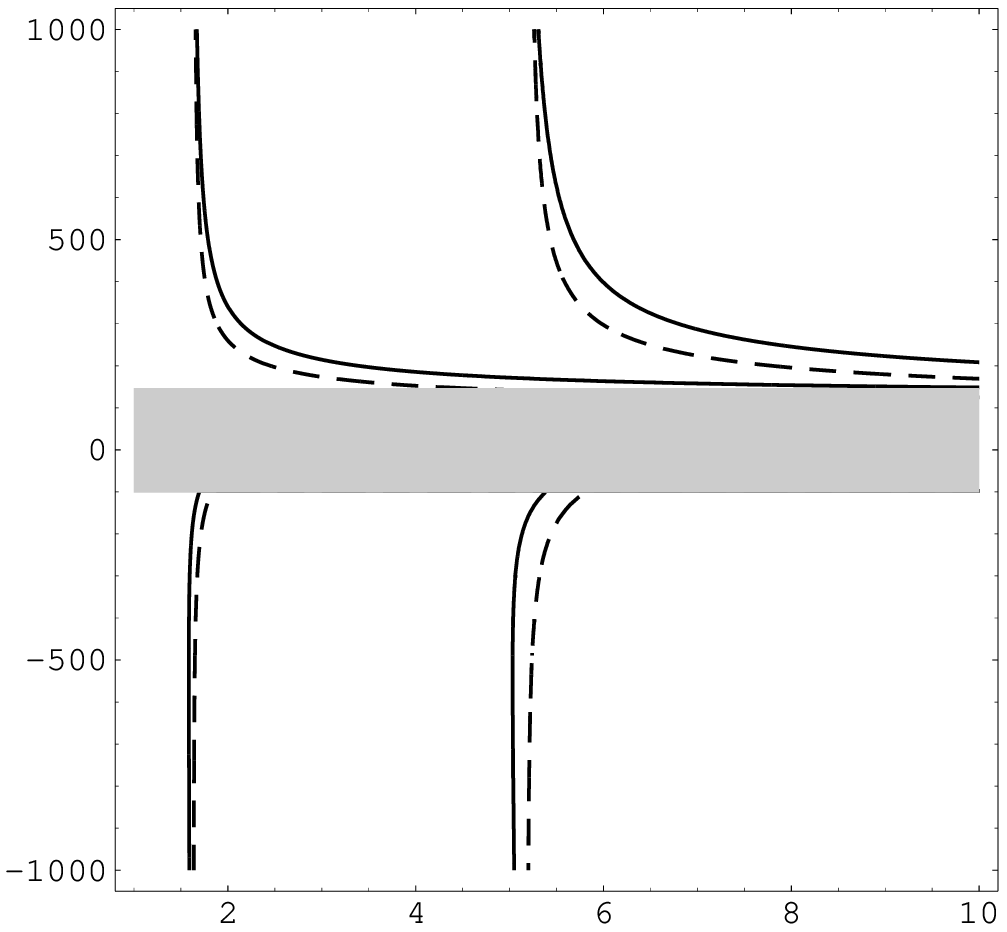,scale=.75}}
\put(0,7.8){$M_2/$GeV}
\put(6.4,.2){$x/$TeV}
\put(3,8.2){(a) NMSSM}
\put(1.6,7.2){\scriptsize $0.99$}
\put(4.2,7.2){\scriptsize $0.999$}

\put(8.3,.5){\epsfig{file=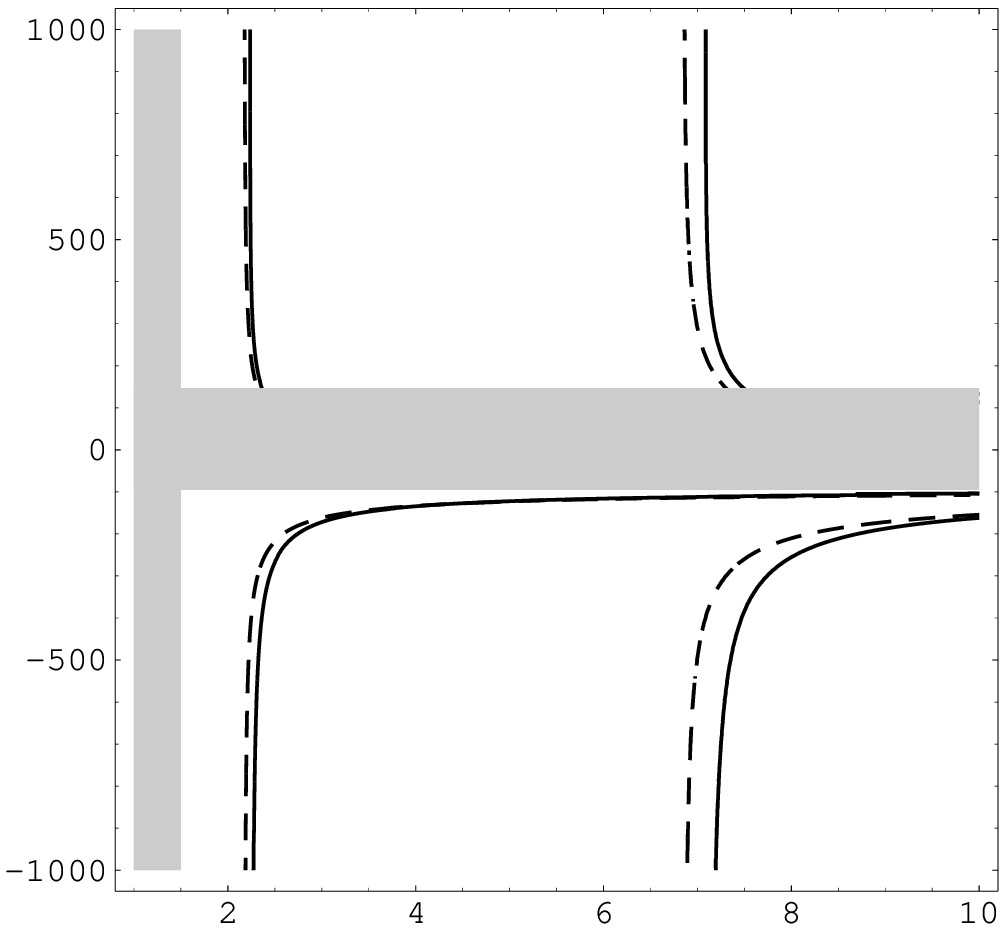,scale=.75}}
\put(8.3,7.8){$M_2/$GeV}
\put(14.7,.2){$x/$TeV}
\put(11.3,8.2){(b) \es\ model}
\put(10.3,7.2){\scriptsize $0.99$}
\put(13.75,7.2){\scriptsize $0.999$}

\end{picture}
\caption{\label{cpnMx} Contour lines of the singlino content of the LSP
$|N_{15}|^2$ and $|N_{16}|^2$ in (a) the NMSSM and (b) the \es\ model,
respectively, with $\tan\beta = 3$ for $\lambda x = 200$~GeV (solid)
and $\lambda x = 400$~GeV (dashed). The LSP mass
$m_{\tilde{\chi}^0_1}=50$~GeV is fixed
by the parameters
$\kappa$ in the NMSSM  and $M'>0$ in the \es\ model. 
The bright
shaded area marks the parameter space experimentally excluded by
the search 
for neutralinos \cite{lepbounds} and new gauge bosons \cite{abe}.}
\end{figure}

Fig.~\ref{cpnMx} shows the contour lines of the singlino content of the LSP
in the NMSSM and the \es\ model 
$|N_{15}|^2$ and $|N_{16}|^2$, respectively,
in the $M_2$-$x$ parameter space for 
$\tan\beta = 3$ and two values $\lambda x = 200$ and 400~GeV.
The mass of the LSP 
$m_{\tilde{\chi}^0_1}=50$~GeV is fixed by the  
parameter $\kappa$ in the NMSSM or $M'$ in the \es\ model.
While the singlino content of the LSP significantly increases
with $x$ it depends only weakly on the other
parameters $M_2$ and $\lambda x $.
In Fig.~\ref{cpnMx}(b) the sign of $M'$ is chosen positive.
Then the contour lines for positive $M_2$ in the NMSSM correspond to those 
with negative
$M_2$ in the \es\ model and vice versa 
because of the relative sign between the submatrices of the minimal and
exotic neutralinos in the neutralino mass matrices 
which is reflected in the signs of the mass eigenvalues of 
$\tilde{\chi}^0_1$ and $\tilde{\chi}^0_2$.
They have the same sign
for positive $M_2$ in the NMSSM, but for negative $M_2$ in the \es\
model.
For $M'<0$ in the \es\ model the mass eigenvalue of the
$\tilde{\chi}^0_1$ flips sign and thus becomes the same as in the
NMSSM, so that in
this case the contour lines of the singlino content of the LSP are
similar to those in the NMSSM.

For values of $x<2$~TeV 
the singlino content of the
LSP is always smaller than about $0.99$ in the whole parameter space
in both models.
For $x=\mathcal{O}(10~\mathrm{TeV})$ it reaches values of about
$0.9997$ in the NMSSM and $0.9995$ in the \es\ model.
In this $x$ range 
the decay widths of the $\tilde{\chi}^0_2$ are typically  
suppressed by a factor  $10^{-2}$ to $10^{-3}$ compared to the MSSM
and so are still too large to yield
displaced vertices.

Obviously, the singlet vacuum expectation value $x$ is the crucial
parameter in order to increase the singlino component of
$\tilde{\chi}^0_1$ and generate displaced vertices of the
$\tilde{\chi}^0_2$. Therefore we discuss in the following the dependence of 
the $\tilde{\chi}^0_1$ singlino content and the $\tilde{\chi}^0_2$ decay width
from the parameter $x$ in four representative supersymmetric scenarios
presented in Table~\ref{scentab}. In all scenarios the lightest neutralino
is singlino dominated with a mass fixed by the parameters $\kappa$ (NMSSM)
or $M'$ (\es\ model). In order to study the impact of the neutralino mixing and
masses we choose scenarios with gaugino and higgsino dominated
$\tilde{\chi}^0_2$ and mass differences 
$m_{\tilde{\chi}^0_2}-m_{\tilde{\chi}^0_1}=10$ GeV and $50$ GeV.

\begin{table}[ht]
\begin{center}
\renewcommand{\arraystretch}{1.3}
\begin{tabular}{|c|c|c|c|c|c|c|}
\hline
Scenario & $M_2/$GeV & $\lambda x$/GeV & $\tan\beta$ & 
$m_{\tilde{\chi}_1^0}/$GeV & $m_{\tilde{\chi}_2^0}/$GeV &
$\tilde{\chi}_2^0$ mixing type\\ 
\hline\hline
G50 & 211 & 400 & 3 & 50 & 100 & gaugino \\ \hline
G90 & 211 & 400 & 3 & 90 & 100 & gaugino \\ \hline
H50 & $-400$ & 107 & 3 & 50 & 100 & higgsino \\ \hline
H90 & $-400$ & 107 & 3 & 90 & 100 & higgsino \\ \hline
\end{tabular}
\end{center}
\caption{\label{scentab}Parameters of the supersymmetric models in
  representative scenarios with a singlino dominated 
  lightest neutralino. The mass of the LSP is fixed by the parameters
  $\kappa$ (NMSSM) and $M'$ (\es\ model).}
\end{table}

\begin{figure}[p]
\begin{picture}(16,17.7)
\put(0,12.5){\epsfig{file=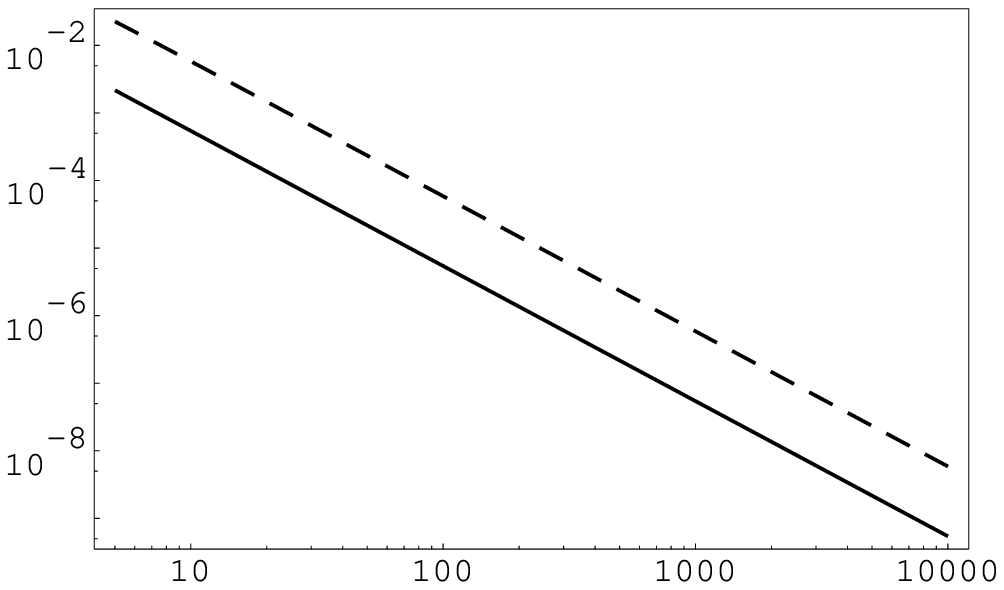,scale=.75}}
\put(6.4,12.1){$x/$TeV}
\put(0,17.1){$1-|N_{15}|^2$}
\put(3,17.3){(a) NMSSM}
\put(4.6,16.55){\small $\tilde{\chi}^0_2$ gaugino-like}
\put(3.2,15.8){\small$m_{\tilde{\chi}^0_1} = 90$ GeV}
\put(1.5,14.3){\small$m_{\tilde{\chi}^0_1} = 50$ GeV}

\put(8.3,12.5){\epsfig{file=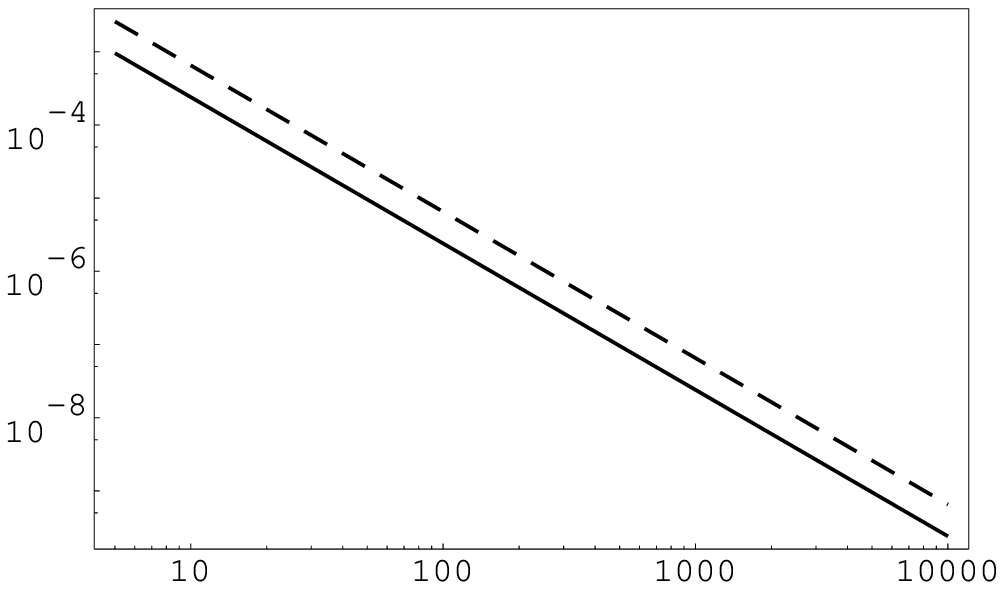,scale=.75}}
\put(14.7,12.1){$x/$TeV}
\put(8.3,17.1){$1-|N_{15}|^2$}
\put(11.3,17.3){(b) NMSSM}
\put(12.9,16.55){\small $\tilde{\chi}^0_2$ higgsino-like}
\put(11.3,15.8){\small$m_{\tilde{\chi}^0_1} = 90$ GeV}
\put(9.8,14.4){\small$m_{\tilde{\chi}^0_1} = 50$ GeV}

\put(0,6.5){\epsfig{file=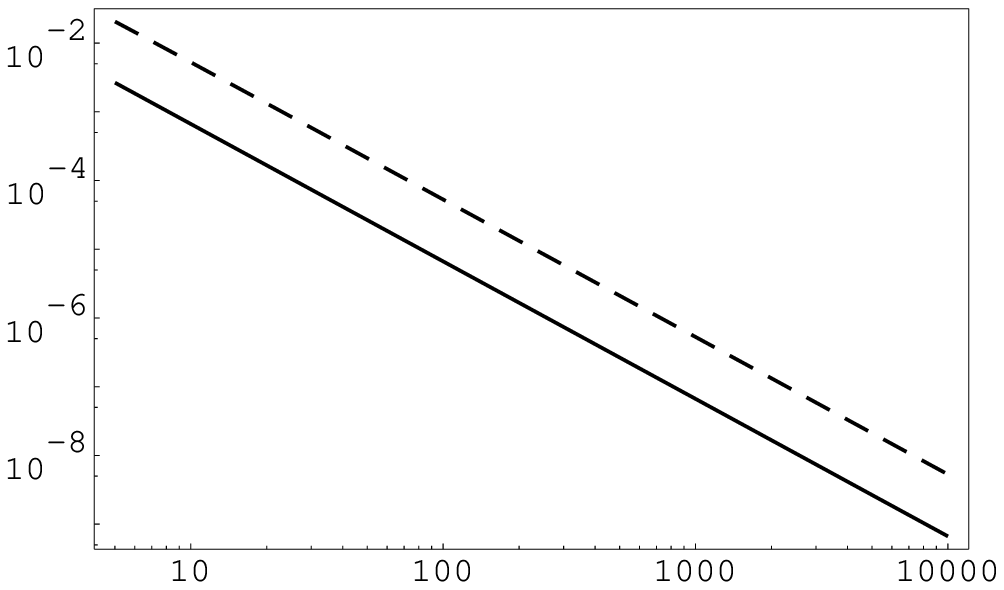,scale=.75}}
\put(6.4,6.1){$x/$TeV}
\put(0,11.1){$1-|N_{16}|^2$}
\put(3,11.3){(c) \es, $M'<0$}
\put(4.6,10.55){\small $\tilde{\chi}^0_2$ gaugino-like}
\put(3.2,9.8){\small$m_{\tilde{\chi}^0_1} = 90$ GeV}
\put(1.5,8.3){\small$m_{\tilde{\chi}^0_1} = 50$ GeV}

\put(8.3,6.5){\epsfig{file=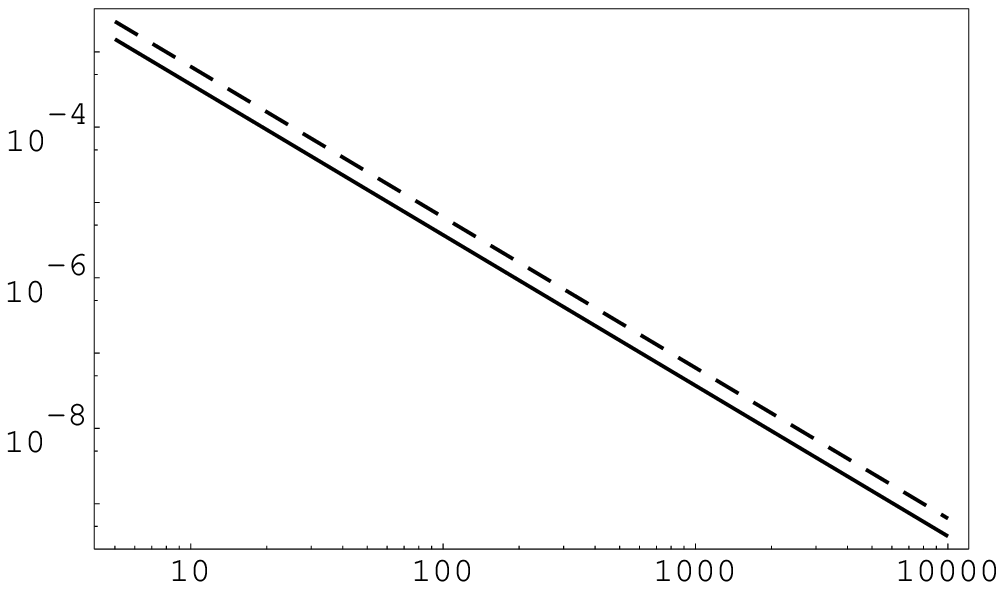,scale=.75}}
\put(14.7,6.1){$x/$TeV}
\put(8.3,11.1){$1-|N_{16}|^2$}
\put(11.3,11.3){(d) \es, $M'<0$}
\put(12.9,10.55){\small $\tilde{\chi}^0_2$ higgsino-like}
\put(11.3,9.8){\small$m_{\tilde{\chi}^0_1} = 90$ GeV}
\put(9.8,8.4){\small$m_{\tilde{\chi}^0_1} = 50$ GeV}

\put(0,.5){\epsfig{file=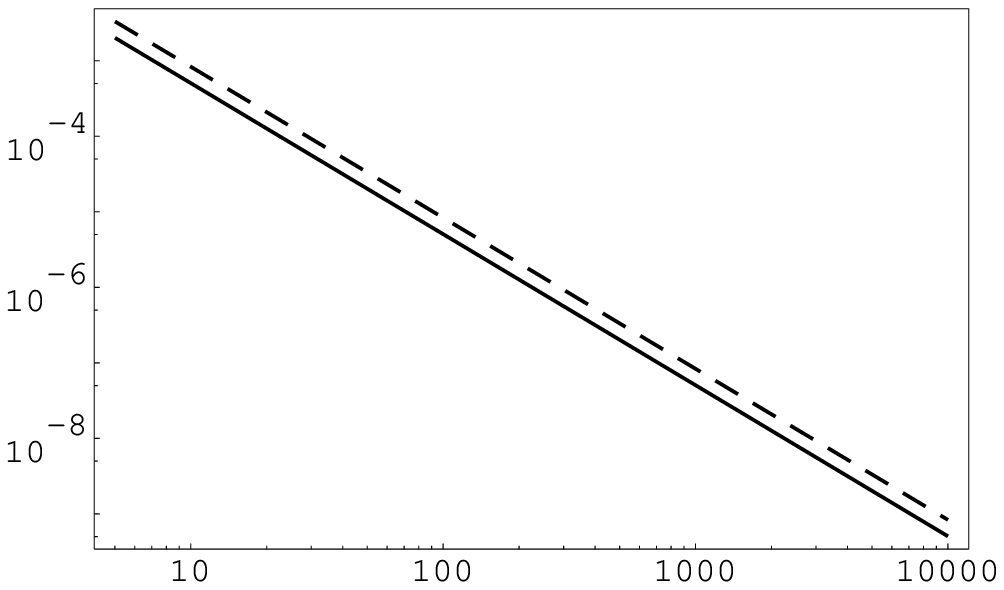,scale=.75}}
\put(6.4,.1){$x/$TeV}
\put(0,5.1){$1-|N_{16}|^2$}
\put(3,5.3){(e) \es, $M'>0$}
\put(4.6,4.55){\small $\tilde{\chi}^0_2$ gaugino-like}
\put(3.1,3.7){\small$m_{\tilde{\chi}^0_1} = 90$ GeV}
\put(1.5,2.5){\small$m_{\tilde{\chi}^0_1} = 50$ GeV}

\put(8.3,.5){\epsfig{file=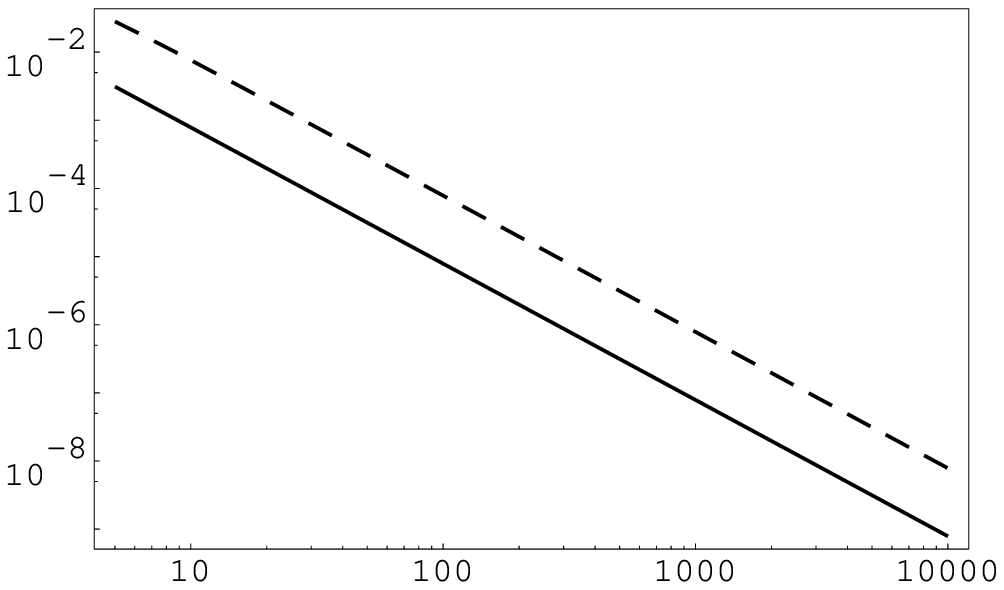,scale=.75}}
\put(14.7,.1){$x/$TeV}
\put(8.3,5.1){$1-|N_{16}|^2$}
\put(11.3,5.3){(f) \es, $M'>0$}
\put(12.9,4.55){\small $\tilde{\chi}^0_2$ higgsino-like}
\put(11.8,3.6){\small$m_{\tilde{\chi}^0_1} = 90$ GeV}
\put(9.7,2.4){\small$m_{\tilde{\chi}^0_1} = 50$ GeV}
\end{picture}
\caption{\label{pn} MSSM content of the LSP with mass $50$~GeV
(solid) and $90$~GeV (dashed) for the scenarios given in
Table~\ref{scentab} in 
(a) the NMSSM with gaugino-like $\tilde{\chi}^0_2$, 
(b) the NMSSM with higgsino-like $\tilde{\chi}^0_2$,
(c) the \es\ model with $M'<0$ and gaugino-like $\tilde{\chi}^0_2$, 
(d) the \es\ model with $M'<0$ and higgsino-like $\tilde{\chi}^0_2$,
(e) the \es\ model with $M'>0$ and gaugino-like $\tilde{\chi}^0_2$ and
(f) the \es\ model with $M'>0$ and higgsino-like $\tilde{\chi}^0_2$.}
\end{figure}

For these scenarios, the MSSM content of the
LSP ($1 - |N_{15}|^2$ and $1 - |N_{16}|^2$, respectively)
is plotted in Fig.~\ref{pn} as a function of $x$. 
In the \es\ model the scenarios (c), (d) differ from (e), (f)
by the sign of the
parameter $M'$ that determines the sign of the mass eigenvalue of the
lightest neutralino because of the see-saw-like mechanism in the
$2\times 2$ submatrix described in Sec.~2.

The MSSM content of the lightest neutralino decreases as $1/x^2$
in very good approximation.
For the larger mass difference between
$\tilde{\chi}^0_1$ and $\tilde{\chi}^0_2$ it 
does not significantly depend on the parameters $M_2$ and $\lambda x$
which determine the mixing character of the $\tilde{\chi}^0_2$ and reaches
values of $10^{-8}$ for singlet vacuum expectation values of about
$3 \times 10^3$ TeV.

Smaller mass differences lead to a larger neutralino mixing and
consequently to a smaller singlino component of the LSP if the
mass eigenvalues of the the light neutralinos have the same sign. 
Then the decoupling of
the submatrices of the MSSM and exotic neutralinos becomes weaker
and the MSSM content of the LSP
is approximately one order of magnitude larger in Fig.~\ref{pn} (a, c, f). 
If, however, the mass eigenvalues have opposite sign, the MSSM content is less
affected by the smaller mass difference in Fig.~\ref{pn} (b, d, e). 
The impact of the MSSM content of the LSP on the decay width of the
NLSP will be discussed in the
next section.

\section{Decay widths}

In this section the decay widths of the second lightest neutralino
$\tilde{\chi}^0_2$ will be discussed
in the scenarios of Table~\ref{scentab}.
Since in the \es\ model the lightest Higgs particle is always a
MSSM-like state
and no light singlet dominated Higgs boson exists for $x>5$~TeV,
two-body decay channels, which could result in much larger decay widths
with displaced vertices appearing at even higher values of $x$, are
kinematically forbidden.
Therefore we only have to consider three-body decays
into leptons, neutrinos or quarks ($\tilde{\chi}^0_2 \to
\tilde{\chi}^0_1 + \ell^+\ell^-$, $\nu\bar{\nu}$, $q\bar{q}$) and
the radiative decay $\tilde{\chi}^0_2 \to \tilde{\chi}^0_1
+ \gamma$.

In the NMSSM, however, light singlet dominated Higgs bosons are not
experimentally excluded. Consequently we discuss in this model also
the case of a light singlet dominated Higgs scalar $S_1$ or 
pseudoscalar $P_1$
and thus include into our analysis the two-body decays
$\tilde{\chi}^0_2 \to \tilde{\chi}^0_1 + S_1 (P_1)$.

The analytical formulae for the neutralino decay width are well
known~\cite{hugonie,suematsu,neuzer}.
For the numerical calculations
we used the 
masses of the scalar leptons and quarks 
given in
Table~\ref{tabsfer}. 
They are motivated by renormalization group equations with the scalar
mass parameter $m_0=135$~GeV and $M_2=300$~GeV in the MSSM and
NMSSM \cite{polchinski}. 
In the \es\ model the same mass values are chosen in order to 
compare the decay widths in NMSSM and \es\ model without sfermion mass 
effects.
Larger or smaller
sfermion masses may obviously reduce or enhance the decay width. 
All qualitative results, however, remain valid.

\begin{table}[ht]
\begin{center}
\renewcommand{\arraystretch}{1.3}
\begin{tabular}{|c|c|c|c|c|c|c|}
\hline
$m_{\tilde{\nu}_L}$/GeV & $m_{\tilde{e}_L}$/GeV & $m_{\tilde{e}_R}$/GeV
& $m_{\tilde{u}_L}$/GeV & $m_{\tilde{u}_R}$/GeV
& $m_{\tilde{d}_L}$/GeV & $m_{\tilde{d}_R}$/GeV \\
\hline
295 & 300 & 200 & 994 & 963 & 996 & 963\\
\hline
\end{tabular}
\end{center}
\caption{\label{tabsfer}Masses of the scalar leptons and
quarks used for the numerical calculation of the
neutralino decay width.}
\end{table}

Displaced vertices are expected for decay lengths of the order
$10^{-3}$ -- 1~m, which correspond to decay widths of $10^{-13}$ --
$10^{-16}$~GeV. At even smaller decay widths the particle escapes
detection and cannot be identified.

\begin{figure}[p]
\begin{picture}(16,17.7)
\put(0,12.5){\epsfig{file=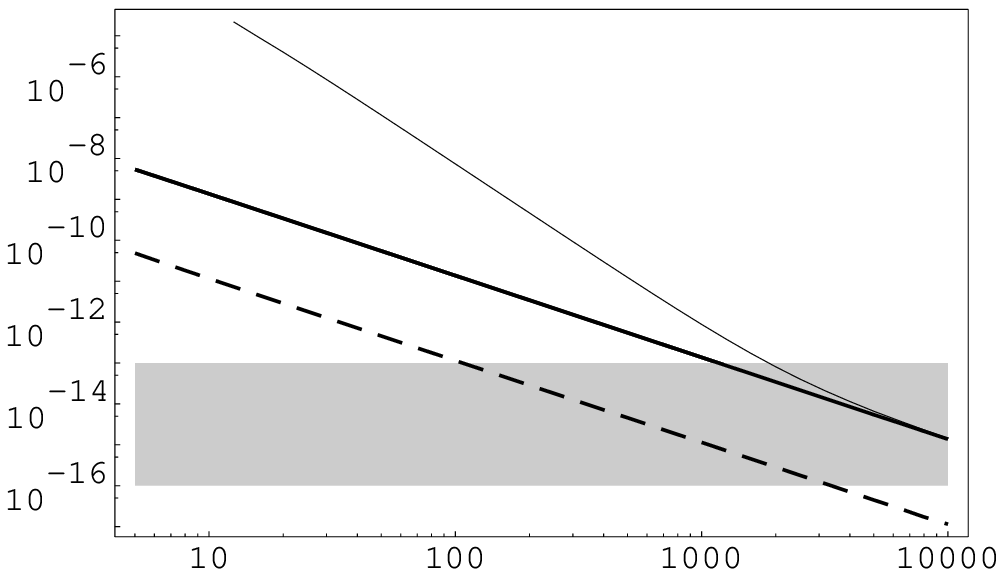,scale=.75}}
\put(6.4,12.1){$x/$TeV}
\put(0,17.1){$\Gamma(\tilde{\chi}^0_2)/$GeV}
\put(3,17.3){(a) NMSSM}
\put(4.6,16.45){\small $\tilde{\chi}^0_2$ gaugino-like}
\put(4.7,15){\small$m_{\tilde{\chi}^0_1} = 50$ GeV}
\put(3.4,13.05){\small$m_{\tilde{\chi}^0_1} = 90$ GeV}

\put(8.3,12.5){\epsfig{file=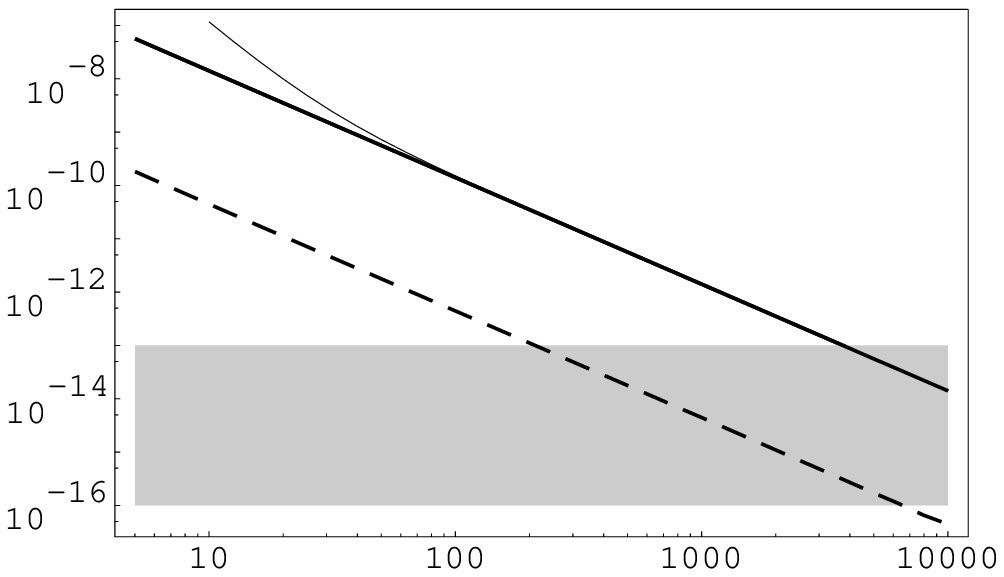,scale=.75}}
\put(14.7,12.1){$x/$TeV}
\put(8.3,17.1){$\Gamma(\tilde{\chi}^0_2)/$GeV}
\put(11.3,17.3){(b) NMSSM}
\put(12.9,16.45){\small $\tilde{\chi}^0_2$ higgsino-like}
\put(13,15.2){\small$m_{\tilde{\chi}^0_1} = 50$ GeV}
\put(9.3,14.2){\small$m_{\tilde{\chi}^0_1} = 90$ GeV}

\put(0,6.5){\epsfig{file=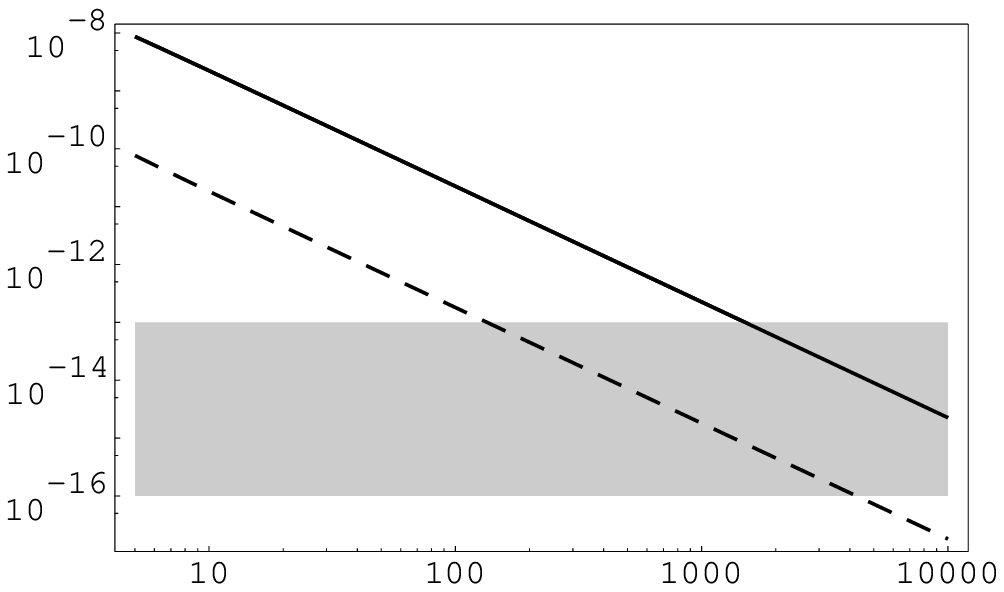,scale=.75}}
\put(6.4,6.1){$x/$TeV}
\put(0,11.2){$\Gamma(\tilde{\chi}^0_2)/$GeV}
\put(3,11.3){(c) \es, $M'<0$}
\put(4.6,10.45){\small $\tilde{\chi}^0_2$ gaugino-like}
\put(3.6,9.7){\small$m_{\tilde{\chi}^0_1} = 50$ GeV}
\put(3.7,7.07){\small$m_{\tilde{\chi}^0_1} = 90$ GeV}

\put(8.3,6.5){\epsfig{file=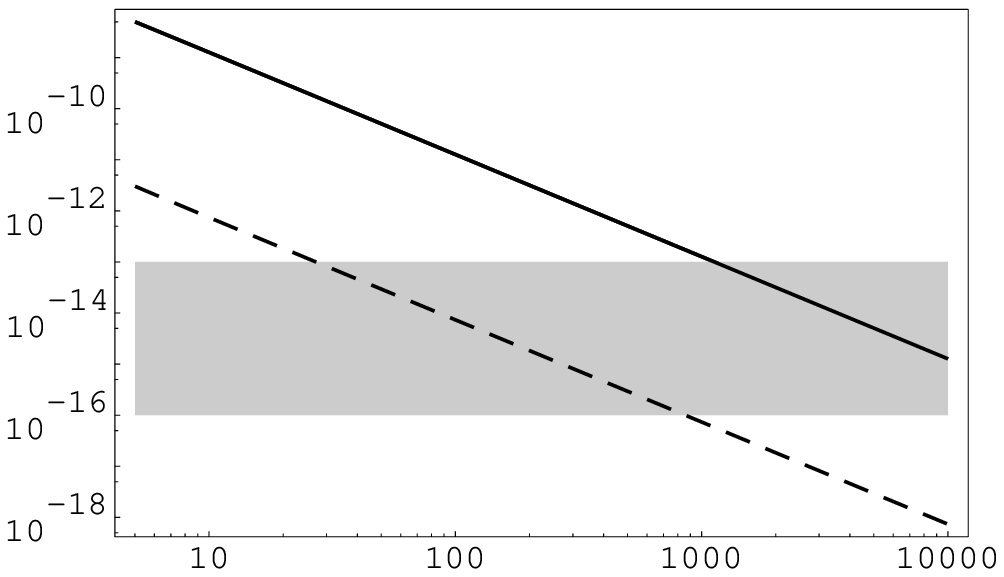,scale=.75}}
\put(14.7,6.1){$x/$TeV}
\put(8.3,11.1){$\Gamma(\tilde{\chi}^0_2)/$GeV}
\put(11.3,11.3){(d) \es, $M'<0$}
\put(12.9,10.45){\small $\tilde{\chi}^0_2$ higgsino-like}
\put(12.1,9.7){\small$m_{\tilde{\chi}^0_1} = 50$ GeV}
\put(11.6,7.2){\small$m_{\tilde{\chi}^0_1} = 90$ GeV}

\put(0,.5){\epsfig{file=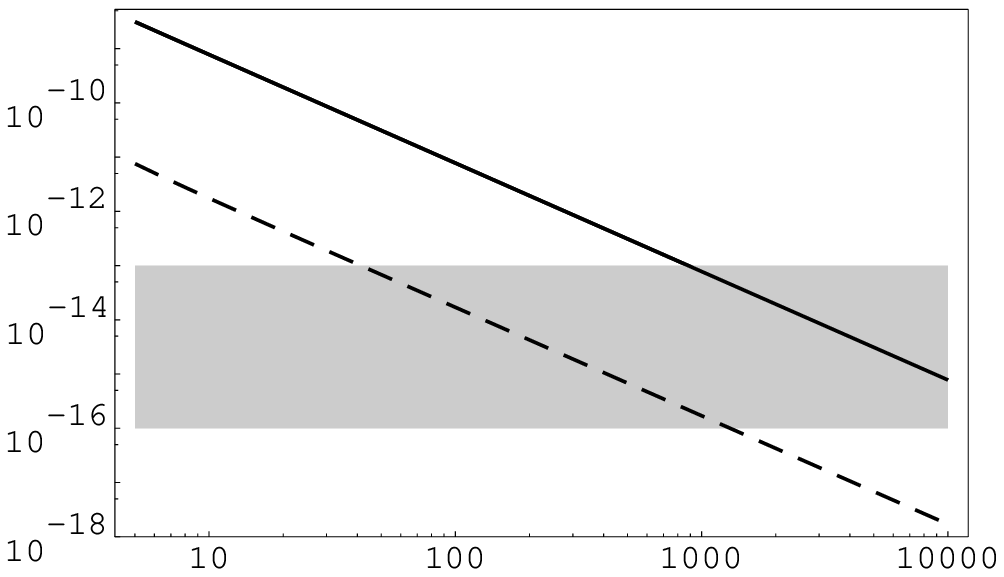,scale=.75}}
\put(6.4,.1){$x/$TeV}
\put(0,5.1){$\Gamma(\tilde{\chi}^0_2)/$GeV}
\put(3,5.3){(e) \es, $M'>0$}
\put(4.6,4.45){\small $\tilde{\chi}^0_2$ gaugino-like}
\put(3.7,3.7){\small$m_{\tilde{\chi}^0_1} = 50$ GeV}
\put(3.3,1.2){\small$m_{\tilde{\chi}^0_1} = 90$ GeV}

\put(8.3,.5){\epsfig{file=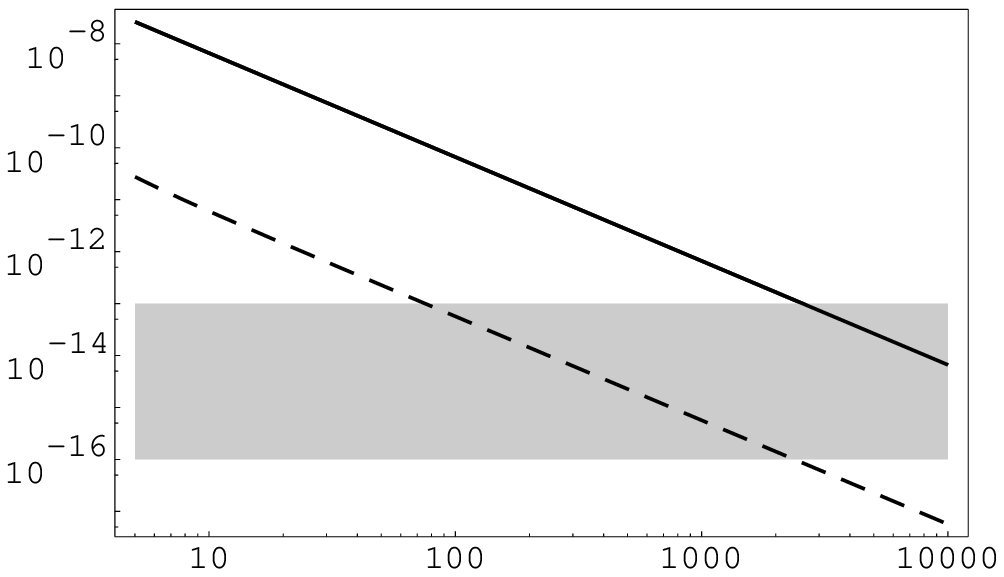,scale=.75}}
\put(14.7,.1){$x/$TeV}
\put(8.3,5.1){$\Gamma(\tilde{\chi}^0_2)/$GeV}
\put(11.3,5.3){(f) \es, $M'>0$}
\put(12.9,4.45){\small $\tilde{\chi}^0_2$ higgsino-like}
\put(12.1,3.7){\small$m_{\tilde{\chi}^0_1} = 50$ GeV}
\put(11.9,1.1){\small$m_{\tilde{\chi}^0_1} = 90$ GeV}
\end{picture}
\caption{\label{decay} Total decay widths of the $\tilde{\chi}^0_2$
with mass $100$~GeV and a mass of the LSP of $50$~GeV
(solid) and $90$~GeV (dashed) for the scenarios given in
Table~\ref{scentab} in 
(a) the NMSSM with gaugino-like $\tilde{\chi}^0_2$, 
(b) the NMSSM with higgsino-like $\tilde{\chi}^0_2$,
(c) the \es\ model with $M'<0$ and gaugino-like $\tilde{\chi}^0_2$, 
(d) the \es\ model with $M'<0$ and higgsino-like $\tilde{\chi}^0_2$,
(e) the \es\ model with $M'>0$ and gaugino-like $\tilde{\chi}^0_2$ and
(f) the \es\ model with $M'>0$ and higgsino-like $\tilde{\chi}^0_2$.
The thin lines in (a) and (b) show the total decay width in the case
of a light singlet-like scalar Higgs of mass 25~GeV in the NMSSM.
The shaded area marks the region of decay widths where displaced
vertices should be visible. Below this area the decaying particle
escapes detection.}
\end{figure}

In Fig.~\ref{decay} the dependence of the total decay width of the
$\tilde{\chi}^0_2$ on the singlet vacuum expectation value $x$ is
shown. 
In all scenarios without open two-body decay channels 
the decay widths decrease approximately as
$1/x^2$ similar as the MSSM content of the LSP studied in Sec.~3.
For the larger mass difference
$m_{\tilde{\chi}^0_2}-m_{\tilde{\chi}^0_1}=50$ GeV
between the decaying $\tilde{\chi}_2^0$
and the LSP the decay widths reach values small
enough for displaced vertices for $x$ between $8\times 10^2$~TeV and
$4\times 10^3$~TeV. 

In the NMSSM scenarios G50 and H50 also the impact of a light
singlet dominated Higgs scalar on the decay width is 
shown in Fig.~\ref{decay}.
As a typical example we set $A_\lambda=0$~GeV and fix
a mass of 25~GeV for the light Higgs scalar by the parameter
$A_\kappa$. 
Then for small $x$ the
two-body decay dominates but decreases approximately as $1/x^4$.
So in G50 slightly larger $x$ values are required in order to observe
displaced vertices 
whereas in H50 the two-body width plays no role for $x>100$~TeV.
Similarly, also the $\tilde{\chi}_2^0$ decay into a light
Higgs pseudoscalar is suppressed at high $x$ values and does
not affect the region of displaced vertices.

The decay widths in the scenarios with smaller
mass difference 
$m_{\tilde{\chi}^0_2}-m_{\tilde{\chi}^0_1}=10$ GeV
are two to three orders of magnitude smaller reaching
the region of displaced vertices already for $x$ between $30$~TeV and
$2\times 10^2$~TeV. They may even become so small that the
$\tilde{\chi}^0_2$ escapes the detector for about $x > 10^3$~TeV.

Although the MSSM content of the LSP becomes larger (Fig.~\ref{pn}),
the decay
widths decrease with smaller mass differences between the NLSP and
the LSP.
Obviously phase space effects clearly outweigh the impact of
the singlino character of the LSP on the decays of
the $\tilde{\chi}^0_2$. Even smaller mass differences, which,
however, make it difficult to detect the the decay products of the
$\tilde{\chi}^0_2$ and thus the displaced vertex,
allow displaced
vertices
already for $x$-values of some TeV.

These results remain valid in all scenarios
with a singlino-like LSP and
mass differences between $\tilde{\chi}^0_1$ and $\tilde{\chi}^0_2$
small enough for two-body decays in $Z$ and MSSM-like
Higgs bosons to be forbidden.
The $\tilde{\chi}^0_2$ decay width is
suppressed by the singlino purity of the LSP which is
governed mainly by the singlet vacuum expectation value $x$
and does only weakly depend on the MSSM parameters
which fix the character of the MSSM-like
$\tilde{\chi}^0_2$.

\section{Conclusion}

Displaced vertices of the NLSP $\tilde{\chi}^0_2$ decaying into a
singlino dominated LSP $\tilde{\chi}^0_1$ may occur
in certain parameter regions of 
the NMSSM and an \es\ model with an additional
singlet Higgs field compared to the MSSM.
They can only be expected for singlet vacuum expectation values 
$x$ in the TeV range significantly above the electroweak symmetry breaking
scale.
For singlet vacuum expectation values of the same order of magnitude as the
vacuum expectation values of the doublet Higgs fields 
displaced $\tilde{\chi}^0_2$ vertices cannot be observed.

In the studied supersymmetric scenarios, minimum singlet vacuum expectation
values between $10^2$ and $10^4$ TeV are required 
for observable displaced $\tilde{\chi}^0_2$ vertices, depending on the mixing
character of $\tilde{\chi}^0_2$ and the mass difference between
$\tilde{\chi}^0_1$ and $\tilde{\chi}^0_2$.
Then the lightest neutralino approaches a singlino state of high purity
with a reduced MSSM content of \mbox{$10^{-4}$ -- $10^{-8}$}.
The $\tilde{\chi}_0^2$ decay into a light singlet-like Higgs
boson, if kinematically allowed, does not significantly affect the 
$x$ region of displaced vertices.
If two-body $\tilde{\chi}^0_2$ decay channels in $Z$ and MSSM-like
Higgs bosons are open, however,
the decay widths are much larger and displaced vertices
appear at considerably higher values of $x$.
For smaller neutralino mass differences $m_{\tilde{\chi}^0_2}-
m_{\tilde{\chi}^0_1}=10$~GeV,
$\tilde{\chi}^0_2$ escapes the detector for
$x$ values larger than about $10^3$ TeV. 

Our results indicate that the appearance of displaced vertices 
in extended supersymmetric models with a singlino dominated
LSP is mainly triggered
by the singlet vacuum expectation value and does not significantly depend
on the supersymmetric parameters of the MSSM.

\section*{Acknowledgment}

We thank X.~Tata and G.~Moortgat-Pick for many helpful comments and
valuable discussions.
S.H. is supported by the Deutsche Forschungsgemeinschaft (DFG) under
contract No.\ HE~3241/1-1.
F.F and H.F. are supported 
by the Deutsche Forschungsgemeinschaft (DFG) under contract No.\
FR~1064/4-1, 
by the Bundesministerium f\"ur
Bildung und Forschung (BMBF) under contract No.
05~HT9WWA~9 and by
the Fonds zur F\"orderung der wissenschaftlichen Forschung of Austria,
project No.~P13139-PHY.

\end{document}